\documentclass[prl,twocolumn,showpacs,showkeys,showpacs,superscriptaddress,10pt,aps]{revtex4-1}
\usepackage{amsmath,amssymb,graphics,graphicx,subfigure,bm}
\usepackage{color}
\usepackage{braket}

\begin{document}

\title{Unexpectedly large electron correlation measured in Auger spectra of ferromagnetic iron thin films: orbital-selected Coulomb and exchange contributions}
\author{R. Gotter}\email{gotter@iom.cnr.it}
\affiliation{CNR-IOM, Istituto Officina dei Materiali, c/o Area Science Park, SS14 km 163.5 I-34149 Basovizza-Trieste, Italy}
\author{A. Verna}
\affiliation{Dipartimento di Scienze, Università degli Studi Roma Tre, Via della Vasca Navale 84, I-00146 Rome, Italy}
\author{M. Sbroscia}\altaffiliation[Current address:]{ Dipartimento di Fisica, Sapienza Universit\'a di Roma, Piazzale Aldo Moro 2, I-00185 Rome, Italy}
\affiliation{Dipartimento di Scienze, Università degli Studi Roma Tre, Via della Vasca Navale 84, I-00146 Rome, Italy}
\author{R. Moroni}
\affiliation{CNR-SPIN, Corso Perrone 24, I-16152, Genova, Italy}
\author{F. Bisio}
\affiliation{CNR-SPIN, Corso Perrone 24, I-16152, Genova, Italy}
\author{S.~Iacobucci}
\affiliation{Dipartimento di Scienze, Università degli Studi Roma Tre, Via della Vasca Navale 84, I-00146 Rome, Italy}
\affiliation{CNR-ISM, Via Fosso del Cavaliere 100, 00133 Roma, Italy}
\author{F. Offi}
\affiliation{Dipartimento di Scienze, Università degli Studi Roma Tre, Via della Vasca Navale 84, I-00146 Rome, Italy}
\author{S. R. Vaidya}\altaffiliation[Current address: ]{Department of Physics, Southern Connecticut State University, 501 Crescent St, New Haven, CT 06515, USA}
\affiliation{CNR-IOM, Istituto Officina dei Materiali, c/o Area Science Park, SS14 km 163.5 I-34149 Basovizza-Trieste, Italy}
\author{A. Ruocco}
\affiliation{Dipartimento di Scienze, Università degli Studi Roma Tre, Via della Vasca Navale 84, I-00146 Rome, Italy}
\author{G. Stefani}
\affiliation{Dipartimento di Scienze, Università degli Studi Roma Tre, Via della Vasca Navale 84, I-00146 Rome, Italy}
\affiliation{CNR-ISM, Via Fosso del Cavaliere 100, 00133 Roma, Italy}

\begin{abstract}
A set of electron-correlation energies as large as 10 eV have been measured for a magnetic 2ML Fe film deposited on Ag(001). By exploiting the spin selectivity in angle-resolved Auger-photoelectron coincidence spectroscopy and the Cini-Sawatzky theory, the core-valence-valence Auger spectrum of a spin-polarized system have been resolved: correlation energies have been determined for each individual combination of the two holes created in the four sub-bands involved in the decay: majority and minority spin, as well as $e_g$ and $t_{2g}$. The energy difference between final states with parallel and antiparallel spin of the two emitted electrons is ascribed to the spin-flip energy for the final ion state, thus disentangling the contributions of Coulomb and exchange interactions.
\end{abstract}
\maketitle

The extraordinary macroscopic properties of technologically relevant materials exploiting quantum effects, including magnetism itself, are mainly determined by the relevance of electron-electron correlations.
The development of novel magnetic and spintronic devices, nowadays evolving more and more at the nanoscale and at the interface level, requires an accurate comprehension of the strongly correlated nature of $d$ or $f$ electron shells, in terms of local and non-local Coulomb and exchange interactions, which depend on energy, orbital, momentum and spin degrees of freedom~\cite{Tusche2018, Steiner1992, SanchezBarriga2009, DaPieve2016a}. 
Density functional theory (DFT)~\cite{KohnSham1965} and computational methods referred as ``Beyond-DFT'' \cite{Mandal2019} provide accurate descriptions of the ground state of moderately-correlated materials while strongly-correlated materials and their excited states still pose significant challenges~\cite{Himmetoglu2014}. These shortcomings are mitigated within Dynamical Mean Field Theory adding, as a tuning parameter, an Hubbard interaction energy $U$~\cite{Hubbard1963, Himmetoglu2014} and an ad-hoc spin-spin interaction term $J$ whenever the magnetic properties have to be taken into account~\cite{Biermann2005}. While spin-resolved photoemission can provide a direct measure of $J$, a wide spectrum of $U$ values is reported in the literature even for the paradigmatic case of iron, whose behavior is at the border between localized and itinerant regimes~\cite{Antonides1977,Steiner1992,Pou2002,SanchezBarriga2009,Katanin2010}. The microscopic understanding of ferromagnetism as a consequence of electron correlations, which changes according to dimensionality, remains a challenging task~\cite{Sthor2006}.
The measurement of the correlation energy is an elusive task for most conventional spectroscopies, since their spectral functions are determined by single-quasi-particle contributions. This is however not the case for core-valence-valence (CVV) Auger spectra, which are sensitive to correlations because of the creation of two interacting holes in the final state and due to the short range of the two-body Coulomb operator acting on many-particle wavefunctions. CVV Auger spectra are described by a two-particle density of states in the valence band (VB), that corresponds to the self-convolution of the independent-particle density of states (SCDOS) only in the absence of correlations~\cite{Weightman1982}.
From the pioneering theory proposed by Cini and Sawatzky (CS)~\cite{Cini1977, Sawatzky1977}, up to the most recent spectral-density-approach (SDA) developed by Nolting and coworkers~\cite{Nolting1991}, an effective electron correlation  $U_{eff}$ is understood to determine the Auger line-shape~\cite{Sawatzky1980, Potthoff1995,Sarma1992}. As $U_{eff}$ becomes comparable to the valence band width $W$, the Auger line-shape changes from purely band-like to atomic-like due to the presence of resonant two-hole final states. Over time, theories have been successfully extended from closed to partially filled bands~\cite{Treglia1981, Verdozzi2001, Sarma1998, Potthoff1993}. In such a perspective, $U_{eff}$ embeds several electron correlation contributions, beyond the on-site Coulomb interaction $U$, thus including the exchange interaction $J$, as well as off-site Coulomb interaction, spin-orbit coupling and dynamical screening effects not included in the CS theory initially formulated for closed bands~\cite{Verdozzi2001}.
$U_{eff}$ can assume different values for different two-hole final states, as found for the weakly correlated CVV Auger spectrum of graphite~\cite{Houston1986}, as well as for the M$_{4,5}$VV spectra of Ag~\cite{Arena2001} and Pd~\cite{Butterfield2002}. Also Auger spectra of spin polarized systems are expected to exhibit different $U_{eff}$ depending on the spin and the band of the electrons involved in the decay process~\cite{Durr1997, Schumann2010, Gotter2012, Wegner2000}.
Ferromagnetic compounds display Coulomb and exchange interactions whose combined action is not negligible with respect to $W$~\cite{Steiner1992, Santoni1991}. Nevertheless, in contrast with the correlated nature of their magnetic properties, ferromagnetic materials exhibit almost band-like Auger spectra, thus hampering the capability to extract information on the correlation of the two-hole final state~\cite{Yin1977, Lund1997}, or even leading to the conclusion that electron correlation is irrelevant~\cite{Antonides1977, Sinkovic1995}.
Angle resolved - Auger photoelectron coincidence spectroscopy (AR-APECS) overcomes this deadlock by combining the selectivity on individual core-hole states \cite{Jiang2001} with the sensitivity to the spin of the Auger final state ~\cite{Gotter2007}, thus allowing correlated final states to be unveiled ~\cite{Gotter2005, Cini2011, Gotter2009, Gotter2011, Gotter2013}.

In this Letter we investigate the role of electron correlation in Auger spectra of a ferromagnetic Fe thin film grown on Ag(001). Due to the non-overlapping $d$ bands of Fe and Ag and to the small population of the Ag $sp$ bands, Fe/Ag(001) is a close approximation to a free standing 2D Fe film ~\cite{Bluegel1992}.
The AR-APECS investigation enables to disentangle features of the Auger spectrum that are originated from electron correlation and the $U_{eff}$ acting on individual pairing of the final state holes, is singled out by the help of the CS model.

\begin{figure}
    \centering
    \includegraphics[width=\columnwidth]{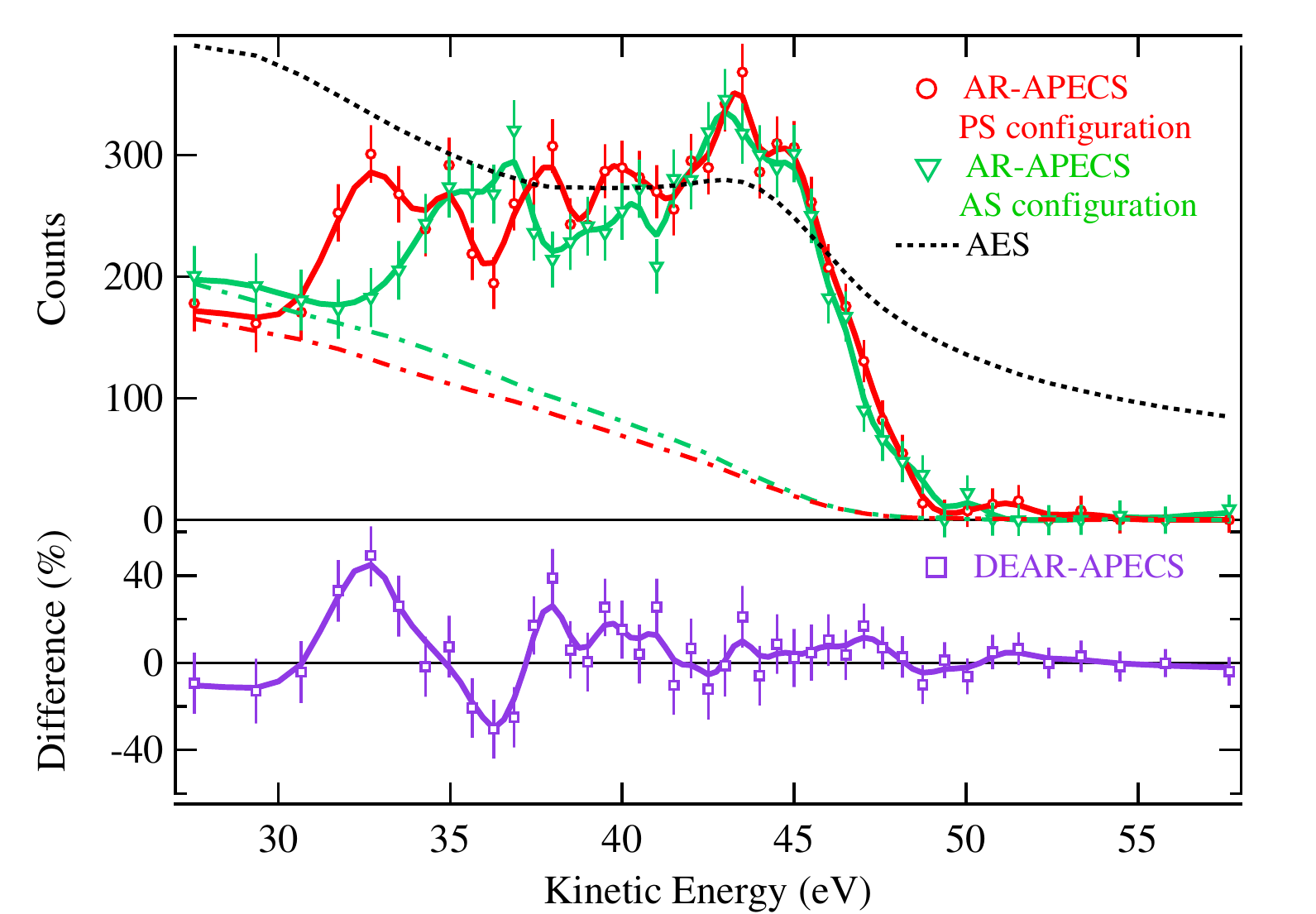}
    \caption{In the upper panel AR-APECS spectra of 2 ML Fe/Ag(001) as measured in the PS (red open circles) and in the AS (green open triangles) configurations are shown together with their estimated integral background (dashed-dotted lines); the continuous lines are guides for the eye. The black dotted line is the conventional Auger electron spectrum (AES), simultaneously collected during the coincidence measurement. In the lower panel the dichroic effect in AR-APECS (DEAR-APECS) (violet open squares) is defined as the difference between PS and AS spectra divided by the semi-sum averaged over the spectrum energy interval (from 30 to 50 eV); the continuous line is a guide for the eyes.}
    \label{fig1}
\end{figure}

\begin{figure}
    \centering
    \includegraphics[width=\columnwidth]{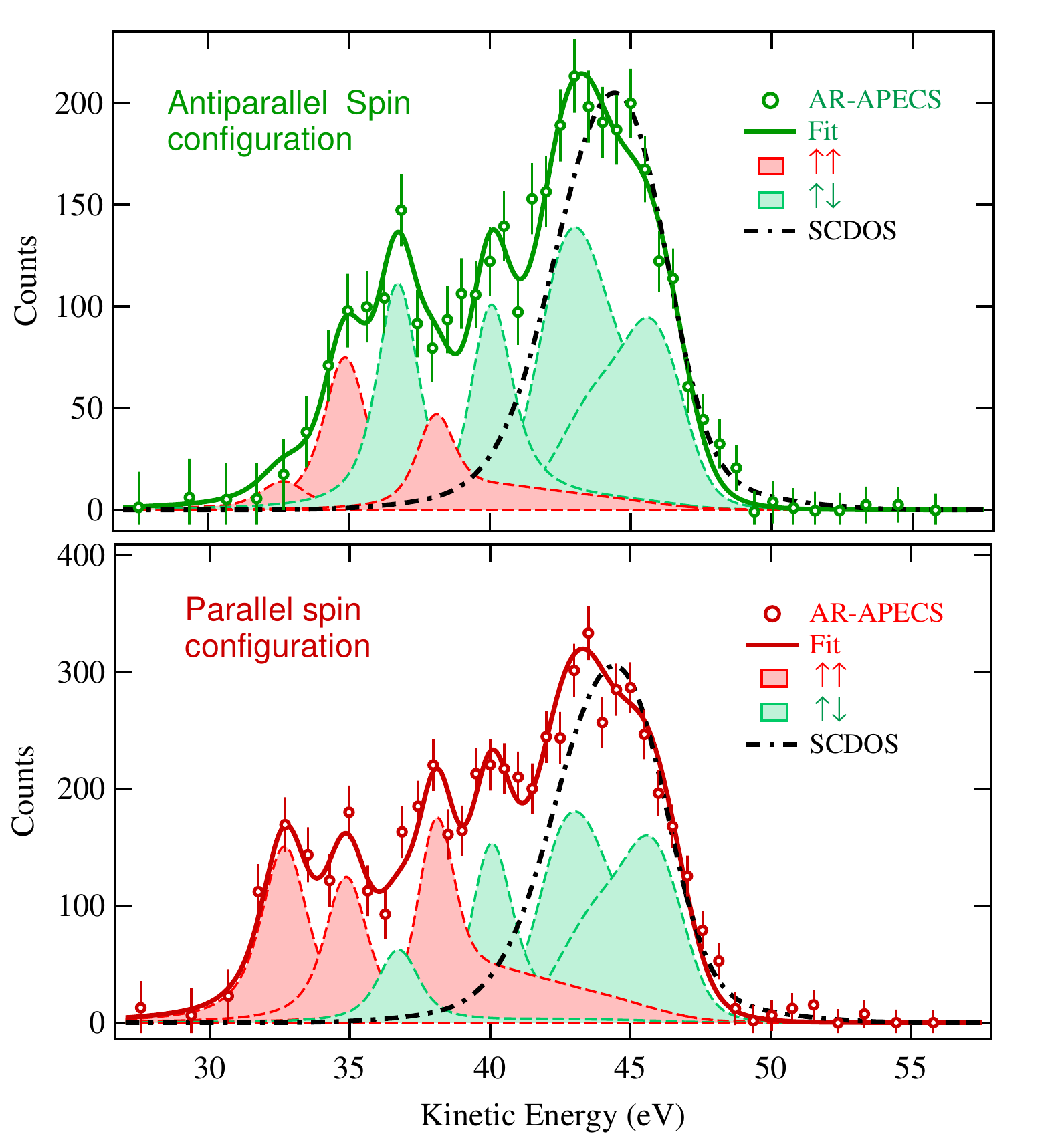}
    \caption{Fe M$_3$VV AR-APECS spectra after inelastic background subtraction in AS (top panel, empty green circles) and PS (bottom panel, red empty circles) configurations. In both panels, the dashed-dotted curve is the SCDOS (no electron correlation) and the red (green) peaks are $\uparrow\uparrow$ ($\uparrow\downarrow$) spin components calculated by the Cini formula. The green and red solid lines are the fitting curves to the experimental data for AS and PS configuration, respectively.}
    \label{fig2}
\end{figure}

The reported experiments were carried out at the ALOISA beamline of the ELETTRA synchrotron radiation facility (Basovizza -Trieste, Italy). The experimental setup is discussed in detail elsewhere~\cite{Gotter2001, Sbroscia2019}.
A 2 mono-layers (ML) thick Fe film was grown at a pressure of 5$\times$10$^{-8}$ Pa, by electron beam assisted evaporation onto the Ag(001) substrate at RT, prepared with standard surface science procedures: 1 KeV Ar$^+$ sputtering and annealing at 750 K. The number of layers was established by the oscillations of the reflection high-energy electron diffraction (RHEED) specular beam intensity~\cite{Luth2010}. The film, kept at 170 K, well below its Curie temperature~\cite{Lang2006}, is ferromagnetic and magnetized out-of-plane~\cite{Schaller1999}. A new sample was prepared every 12 h of beam exposure to prevent oxidation.
A monochromatic, linearly polarized  beam of 253 eV photons impinged onto the sample surface at a grazing angle of about 6$^{\circ}$, with the surface normal lying in the plane defined by the photon polarization $\bm\varepsilon$ (the quantization axis) and the momentum $\mathbf{k}$ vectors. Auger- and photo-electron pairs were selected in energy and detected in coincidence within the solid angles (4$^{\circ}$ opening) defined by the electron analyzers. Three analyzers placed in the $\bm\varepsilon\mathbf{k}$ plane collected Fe 3$p$ core photoelectrons at 0$^{\circ}$ and $\pm$36$^{\circ}$ polar angle with respect to $\bm\varepsilon$, with an energy resolution of 3.2 eV; they were detuned by 1.5 eV at higher kinetic energy with respect to the 3$p$ maximum photoemission intensity~\cite{Gotter2009} for collecting mainly the three photoemission lines closely packed at the high kinetic energy side of the $3p$ sextet~\cite{Tamura1994,Henk1999}. Auger electrons resulting from the  Fe M$_{3}$M$_{4,5}$M$_{4,5}$ super-Coster-Kronig transition have been collected at an angle of 38$^{\circ}$ off the $\bm\varepsilon\mathbf{k}$ plane via a multichannel analyzer with an energy resolution of 1 eV.
The electron pairs detected at the three different photoelectron emission angles allow to exploit the dichroic effect in angle resolved APECS (DEAR-APECS) thereby providing a moderate selection of the final state spin~\cite{Gotter2005}. In analogy to previous AR-APECS experiments carried out with the same geometries~\cite{Gotter2013}, in the present experiments the analyzer pair selecting photoelectrons emitted along $\bm\varepsilon$ will be termed as antiparallel-spin (AS) configuration, while the pairs with photoelectrons detected at $\pm$36$^{\circ}$ apart from $\bm\varepsilon$, will be termed as parallel-spin (PS), as they detect electron pairs with predominantly antiparallel and parallel spins, respectively. The coincidence count rate for these experiments was of the order of 1.3 10$^{-1}$ counts per second, thus, to achieve a good statistics, nearly 40 h of integration time was required.
In the upper panel of Fig.~\ref{fig1} the AR-APECS spectra as measured in AS and PS configurations are shown together with the conventional Auger (AES) measured under identical experimental conditions. The AR-APECS spectra show a rich multiplicity of narrow transitions spread across an energy interval larger than twice the band width ($W$), i.e. well beyond the interval from 49 to 40 eV allowed by energy conservation if electronic correlation were not effective~\cite{Schroeder1985}.
Not even spin-resolved Auger investigations~\cite{Sinkovic1995, Schroeder1985, Allenspach1987} disclosed these manifold structures, because detecting the spin of only one electron (the Auger electron) does not allow a selectivity on the two-hole final state. For instance, the majority spin ($\uparrow$) Auger intensity was represented by terms proportional to the self-convolution of the majority-spin DOS ($\uparrow\uparrow$ terms) plus terms proportional to the convolution of the majority- with the minority-spin DOS ($\uparrow\downarrow$ terms)~\cite{Sinkovic1995}, that is by a mix of terms having different total spin $S$, used to identify the final states in the LS coupling.
The DEAR-APECS~\cite{Gotter2009} reported in the lower panel of Fig.~\ref{fig1}, that is the difference between the PS and the AS spectra,  is more pronounced in the low kinetic energy region (from 30 to 37 eV). The CS model foresees an Auger lineshape consisting of a manifold of closely-spaced sub-bands contributions on the high energy side of the spectrum, accompanied by sharper features at lower kinetic energy due to resonant two-hole states~\cite{Cini1977, Sawatzky1977, Verdozzi2001}. A previous AR-APECS investigation on the Fe MVV spectrum of a 3ML film of Fe/Cu(001), although performed with a moderate 2 eV energy resolution, has associated the low-energy region of the spectrum to resonant two-hole final states with a single average correlation energy of 2.7 eV, estimated by applying the Cini formula~\cite[formula (14)]{Cini1977} to a DFT computed DOS of a Fe impurity in a Cu jellium~\cite{Gotter2012}.
After the removal of an integral background due to energy losses suffered by the Auger electrons ~\cite{Seah2001}, the AS and PS spectra reported in Fig.~\ref{fig2} unravel a manifold of features that are interpreted as individual two-hole correlation resonances. The relative maxima at 32.4, 34.9 and 37.6 eV, which are prominent in the PS configuration and strongly reduced in the AS one, can be ascribed to the parallel spin of the two emitted electrons; vice versa, the antiparallel spin character can be attributed to the structure at 36.6 and 40.3 eV, which are dominant in the AS configuration.
The data analysis reported in the following, builds on the Fe DOS calculated in DFT-LSDA by Rhee~\cite{Rhee2005} for a 3 ML Fe ultrathin film on Ag(100) and for bulk Fe, the latter being in very good agreement with a recent (DFT-LSDA) calculation~\cite{DaPieve2016a}. The 2 ML-thick film here investigated has been modelled using Rhee's DOS of the surface (top) and interface (bottom) layers; this latter weighted by the effective mean free path of the emitted electron pair~\cite{Werner2005}.
The dashed-dotted line in Fig.~\ref{fig2} is the self(mutual)-convolution of the density of states (SCDOS) summed over all the spin-resolved occupied $e_g$ and $t_{2g}$ bands and broadened by the experimental resolution; it describes the band-like Auger line-shape in absence of electron correlation, and it accounts for the measured intensity only from the onset up to the maximum of the AR-APECS intensity, without fitting any of the sharp features.
If the atomic multiplet of Fe is taken into account (see figures and ref.~\onlinecite{supplemento}), the contribution of 16 multiplet terms over an energy interval of 13 eV does not cover the full width of the measured AR-APECS spectra; not even if to such an atomic multiplet, the CS model is applied, as successfully done in the case of Pd and Ag MVV Auger spectra~\cite{Butterfield2002, Arena2001,Weightman1982}.
Therefore the CS model was applied to each possible pairing of the individual spin and orbital components of the theoretical DOS as derived by Rhee~\cite{Rhee2005} and a set of $U_{eff}$ values were used as free parameters. In particular the Cini formula~\cite{Cini1977} provide a formulation of the two-particle spectral density as a simple functional of one-particle DOS, suitable to be implemented in a fitting procedure. A least square fitting procedure has been simultaneously applied to both PS and AS AR-APECS spectra with a tentative assignment of the manifold features made on the basis of the following considerations:
{\it I}. according to the Hund's rule, final states of the ion left behind corresponding to the emission of two ($\uparrow\uparrow$) electrons have a lower total spin, so a higher binding energy has to be expected with respect to final states corresponding to the antiparallel spin ($\uparrow\downarrow$) of the two emitted electrons;
{\it II}. the above assessment also comply with the fact that majority spin bands are almost filled bands while minority ones are open, hence higher $U_{eff}$ are expected for final states where both holes are created in the majority spin bands  ($\uparrow\uparrow$);
{\it III}. due to the small hybridization between $e_g$ and $t_{2g}$ bands, $e_g$ behaves like a Luttinger (localized) electron liquid while $t_{2g}$ behaves like a Fermi (itinerant) liquid~\cite{Katanin2010}; hence Auger transitions involving $e_g$ states should be credited for larger $U_{eff}$;
{\it IV}. the electron correlation between two holes created in the same sub-band is larger with respect to the case when they are created in different sub-bands;
{\it V}. the ($\downarrow\downarrow$) contribution to the Auger intensity is neglected in view of the smallness of the resulting SCDOSs.
Such a rationale, together with the possibility to identify the three ($\uparrow\uparrow$) contributions (red filled peaks in Fig.~\ref{fig2}), that are predominant in the PS configuration, and the four ($\uparrow\downarrow$) contributions (green filled peaks in Fig.~\ref{fig2}), that are relatively more intense in the AS configuration, allowed to set the initial guess of $U_{eff}$ values in the fitting procedure.
A chi-square likelihood test of the best fit procedure provided the set of $U_{eff}$ parameters shown in Table~\ref{tab1} and the resulting Auger line-shapes are the continuous thick lines in Fig.~\ref{fig2}. 

\begin{table*}
\begin{tabular}{|l|c|c|c|c|c|c|c|c|}
\hline
Hole paring	& $e_g^\uparrow e_g^\uparrow$ & $t_{2g}^\uparrow t_{2g}^\uparrow$ &	$e_g^\uparrow e_g^\downarrow$ &	$e_g^\uparrow t_{2g}^\uparrow$ & $t_{2g}^\uparrow t_{2g}^\downarrow$ &	$t_{2g}^\uparrow  e_g^\downarrow$ & $e_g^\uparrow t_{2g}^\downarrow$ & SCDOS\\
\hline
Kinetic Energy [eV $\pm$0.1 eV] & 32.6 & 34.9 & 36.7 & 37.7 & 40.0 & 42.7 & - & 44.4 \\
\hline
$U_{eff}$ [eV $\pm$0.1 eV] & 11.0 & 8.7 & 7.5 & 5.20 & 4.3 & 1.0 & 0.0 & 0.0 \\
\hline
\end{tabular}
\caption{$U_{eff}$ values for different pairings of the two emitted electrons, as obtained from the simultaneous fitting of the Fe M$_{3}$VV AR-APECS spectra in the AS and PS configuration of Fig.~\ref{fig2}. The kinetic energy position of each component is also reported.}
\label{tab1}
\end{table*}
An unexpected spread of $U_{eff}$ values ranges from almost vanishing values (cases $e_g^\uparrow t_{2g}^\downarrow$ and  $t_{2g}^\uparrow e_g^\downarrow$) to figures much larger than any reported observation (cases $e_g^\uparrow e_g^\uparrow$ and $t_{2g}^\uparrow t_{2g}^\uparrow$). It is worth noting that the difference in $U_{eff}$ between transitions involving identical band combinations, but with parallel and antiparallel spin of the two emitted electrons, corresponds to the spin-flip energy of an electron in the doubly ionized final state of the ion left behind. A spin-flip energy is used to define the exchange-splitting in the Stoner model, which, in the case of Fe, amounts to about 2 eV when calculated for the ground states of neutral Fe or measured by photoemission experiments~\cite{Sthor2006}.
Here, the quantities $U_{eff}(t_{2g}^\uparrow t_{2g}^\uparrow) -  U_{eff}(t_{2g}^\uparrow  t_{2g}^\downarrow) =$ 4.4 $\pm$ 0.2 eV, and $U_{eff}(e_g^\uparrow  t_{2g}^\uparrow) -  U_{eff}(e_g^\uparrow  t_{2g}^\downarrow) =$ 5.2 $\pm$ 0.2 eV, can be identified as the energy necessary to flip a $t_{2g}$ electron in two different electronic configurations; the latter value is larger by an amounts of 0.8 $\pm$0.2 eV because the spin-flip is paired with a higher number of $t_{2g}$ electrons.
These $t_{2g}$ spin-flip energies result in turn higher with respect to the ones associated to the $e_g$ orbital, according with a higher $d$-electron occupation number of the $t_{2g}$ sub-bands with respect to $e_g$ ones~\cite{Rhee2005}; in detail $U_{eff}(e_g^\uparrow e_g^\uparrow) - U_{eff}(e_g^\uparrow  e_g^\downarrow) =$ 3.5 $\pm$ 0.2 eV, and $U_{eff}(t_{2g}^\uparrow e_g^\uparrow)  - U_{eff}(t_{2g}^\uparrow e_g^\downarrow) =$ 4.2 $\pm$ 0.2 eV, with the latter value larger by an amount of 0.7 $\pm$0.2 eV because the spin-flip is paired with a higher number of $e_{g}$ electrons. The two values of 0.82 and 0.72 eV, are equal within the experimental uncertainty and provide a final consistency of the assignments of $U_{eff}$ made so far.
The very large values here found for $U_{eff}$ are therefore due to the combined action of the on-site Coulomb interaction and the spin-flip energies, these latter being experimentally singled out for the first time in a CVV Auger spectrum of a spin polarized system.
The capability to reveal itemized experimental values of the electron correlation allows to test more recent theoretical models~\cite{Wegner2000}, that can benefit of the exploitation of multi-band Hamiltonians to better determine the details of electron correlation in magnetic systems, and from which the three-particle Green function, needed to calculate the Auger line-shape, can be developed.
In conclusion, by AR-APECS measurements on an ultrathin ferromagnetic Fe film the CVV Auger spectrum of a spin polarized system has been fully resolved, giving access with unprecedented accuracy to electron correlation effects, due to Coulomb interaction and  spin coupling of the valence holes created in the final state. By exploiting the Cini-Sawatzky theory, a full set of electron correlation energies $U_{eff}$ has been determined. Noticeable differences among the $U_{eff}$ associated to the different pairings of the spin-polarized sub-bands involved in the Auger decay have been found. $U_{eff}$ values, much larger than the ones found in the literature, are responsible for the manifold of two-holes resonances that remained undetected by conventional Auger. The $U_{eff}$ used to describe Auger line-shapes, when acting on a spin polarized system, has been understood as due to two effects: the first is a Coulomb correlation energy, which depends on the orbitals involved in the Auger decay; the second is a spin-flip energy characterizing the energy difference of final states associated to parallel and antiparallel spin of the emitted electrons.

\paragraph{Acknowledgements}
Financial supports through: SIMDALEE2 Marie Sklodowska Curie FP7-PEOPLE-2013-ITN Grant \# 606988, PRIN 2015 NEWLI Protocol 2015CL3APH, and ELETTRA users support are acknowledged. The authors are grateful for the support of the staff of the ALOISA beamline of the ELETTRA synchrotron. One of us (R.G.) is grateful to D.D. Sarma for the fruitful discussion. The authors are deeply indebted to Fabiana Da Pieve for enlightening discussions.

\bibliographystyle{apsrev4-1}
\bibliography{FeAg}

\end{document}